\documentclass[journal,twoside,web]{ieeecolor}
\usepackage{etoolbox}
\makeatletter
\@ifundefined{color@begingroup}%
{\let\color@begingroup\relax
\let\color@endgroup\relax}{}%
\def\fix@ieeecolor@hbox#1{%
\hbox{\color@begingroup#1\color@endgroup}}
\patchcmd\@makecaption{\hbox}{\fix@ieeecolor@hbox}{}{\FAILED}
\patchcmd\@makecaption{\hbox}{\fix@ieeecolor@hbox}{}{\FAILED}
\usepackage{generic}
\usepackage{cite}
\usepackage{amsmath,amssymb,amsfonts}
\usepackage{algorithmic}
\usepackage{graphicx}
\usepackage{textcomp}
\usepackage{multirow}
\usepackage{array}
\usepackage{makecell}
\usepackage{color}
\usepackage{soul}
\usepackage[export]{adjustbox} 
\usepackage{bm}
\usepackage{tabularx}
\usepackage{caption}
\usepackage{booktabs} 
\usepackage{float}

\soulregister\cite7 
\soulregister\citep7 
\soulregister\citet7 
\soulregister\ref7 
\soulregister\pageref7 
\makeatletter
\let\NAT@parse\undefined
\makeatother
\usepackage[colorlinks=true,
           linkcolor=blue,
           anchorcolor=blue,
           citecolor=green,
           urlcolor=blue,
           ]{hyperref}
\def\BibTeX{{\rm B\kern-.05em{\sc i\kern-.025em b}\kern-.08em
    T\kern-.1667em\lower.7ex\hbox{E}\kern-.125emX}}
\begin{document}
\title{BS-LDM: Effective \textbf{B}one \textbf{S}uppression in High-Resolution Chest X-Ray Images with Conditional \textbf{L}atent \textbf{D}iffusion \textbf{M}odels}
\author{Yifei Sun, Zhanghao Chen, Hao Zheng, Wenming Deng, Jin Liu,  Wenwen Min,\\ Ahmed Elazab, Xiang Wan, Changmiao Wang, Ruiquan Ge, \IEEEmembership{Member, IEEE}
\thanks{This work was supported by  the Open Project Program of the State Key Laboratory of CAD\&CG (No. A2410), Zhejiang University, Zhejiang Provincial Natural Science Foundation of China (No. LY21F020017), National Natural Science Foundation of China (No. 61702146, 62076084, U22A2033, U20A20386), Guangdong Basic and Applied Basic Research Foundation (No. 2025A1515011617, 2022A1515110570), Innovation Teams of Youth Innovation in Science and Technology of High Education institutions of Shandong Province (No. 2021KJ088). (Corresponding author: Changmiao Wang, Ruiquan Ge).}
\thanks{Y. Sun, Z. Chen, H. Zheng and R. Ge are with Hangzhou Dianzi University, Hangzhou, China (e-mail:szhsxhsyf@hdu.edu.cn, czh345068@gmail.com, (22320210, gespring)@hdu.edu.cn).}
\thanks{W. Deng is with National Cancer Center/National Clinical Research Center for Cancer/Cancer Hospital\&Shenzhen Hospital, Chinese Academy of Medical Sciences and Peking Union Medical College, Shenzhen, China (email: ming861212@163.com).}
\thanks{J. Liu is with Central South University, Changsha, China (e-mail:liujin06@csu.edu.cn).}
\thanks{W. Min is with Yunnan University, Kunming, China (e-mail:minwenwen@ynu.edu.cn).}
\thanks{A. Elazab is with the School of Biomedical Engineering, Shenzhen University, Shenzhen, China and Computer Science Department, Misr Higher Institute for Commerce and Computers, Mansoura, Egypt (email:ahmed.elazab@yahoo.com).}
\thanks{X. Wan and C. Wang are with Shenzhen Research Institute of Big Data, Shenzhen, China (e-mail:wanxiang@sribd.cn; cmwangalbert@gmail.com).}
}

\maketitle
\begin{abstract}
Lung diseases represent a significant global health challenge, with Chest X-Ray (CXR) being a key diagnostic tool due to its accessibility and affordability. Nonetheless, the detection of pulmonary lesions is often hindered by overlapping bone structures in CXR images, leading to potential misdiagnoses. To address this issue, we develop an end-to-end framework called BS-LDM, designed to effectively suppress bone in high-resolution CXR images. This framework is based on conditional latent diffusion models and incorporates a multi-level hybrid loss-constrained vector-quantized generative adversarial network which is crafted for perceptual compression, ensuring the preservation of details. To further enhance the framework's performance, we utilize offset noise in the forward process, and a temporal adaptive thresholding strategy in the reverse  process. These additions help minimize discrepancies in generating low-frequency information of soft tissue images. Additionally, we have compiled a high-quality bone suppression dataset named SZCH-X-Rays. This dataset includes 818 pairs of high-resolution CXR and soft tissue images collected from our partner hospital. Moreover, we processed 241 data pairs from the JSRT dataset into negative images, which are more commonly used in clinical practice. Our comprehensive experiments and downstream evaluations reveal that BS-LDM excels in bone suppression, underscoring its clinical value. 
Our code is available at \href{https://github.com/diaoquesang/BS-LDM}{https://github.com/diaoquesang/BS-LDM}.
\end{abstract}


\section{Introduction}
\label{sec:introduction}

\IEEEPARstart{L}{ung} disease remains a leading cause of high morbidity and mortality worldwide \cite{pham2021cnn,zhen2025epidemiology}. Chest X-Ray (CXR) is the primary imaging technique for evaluating conditions such as inflammation, tuberculosis, and lung masses owing to its accessibility, cost-effectiveness, and low radiation exposure. However, even skilled radiologists may miss lesions in CXR images. This challenge primarily arises from bone structures overlapping with the lung region, which can obscure crucial details necessary for accurate detection and diagnosis. Research indicates that between 82\% and 95\% of undetected lung cancers are concealed by these bone structures \cite{austin1992missed}. Consequently, bone suppression techniques in CXR images could greatly aid radiologists \cite{li2020high}, and enhance the performance of computer-aided pulmonary lesion detection \cite{wang2025rib}.


Dual-Energy Subtraction (DES) imaging is a technique used to reduce visual clutter in CXR images caused by overlapping bone structures \cite{vock2009dual}. DES radiography uses two X-ray exposures at different energy levels to create two separate radiographs. These are then combined to produce a single image that emphasizes either soft tissue or bone structures. For soft tissue imaging, removing the bone component significantly enhances the clarity of chest radiographs. However, DES imaging requires specialized equipment and results in higher radiation doses, which limits its accessibility and use in developing and low-income countries. This limitation poses challenges in acquiring high-quality DES soft tissue images.

An alternative to DES imaging for bone suppression in CXR images is image processing, which does not require specialized equipment. Traditional image processing approaches treat bone suppression as a regression prediction problem, using supervised methods. For instance, Suzuki \textit{et al.} \cite{suzuki2006image} utilized a multi-task artificial neural network to generate bone images from CXR images, allowing for the subtraction of these bone images to yield soft tissue images. However, the small dataset used limited the generalizability of their approach. Following this, statistical analysis methods were developed to identify and remove bony structures using image features in an unsupervised manner. Simkó \textit{et al.} \cite{simko2009elimination} introduced a clavicle suppression algorithm that produced a bone image from a gradient map, adjusted along the bone boundary. This generated bone image was then subtracted from the CXR images to obtain soft tissue images. Similarly, Juhasz \textit{et al.} \cite{juhasz2010segmentation} applied an active shape model for segmenting anatomical structures in CXR images, effectively suppressing bone shadows. They implemented this model on the JSRT dataset \cite{shiraishi2000development}, which remains the only publicly accessible dataset of its kind. However, statistical-based methods generally require precise segmentation and boundary annotations, which often lack rich semantic information regarding bony structures.

With advancements in neural networks, deep learning-based algorithms have increasingly been used for bone suppression in CXR images, generating soft tissue images effectively. These methods typically approach bone suppression as either an end-to-end image denoising task or focus on predicting the bone residual. Generally, end-to-end image denoising yields better bone suppression, but it may compromise detail preservation due to model limitations. Conversely, the bone residual prediction approach often produces impressive results in the generation of lesion and texture, although it struggles to completely remove bones. Yang \textit{et al.} \cite{yang2017cascade} presented a cascaded multi-scale Convolutional Neural Network (CNN) trained within the gradient domain of CXR images for this purpose. While the model demonstrated impressive performance, it failed to maintain high levels of perceptual and structural integrity. Similarly, Gusarev \textit{et al.} \cite{gusarev2017deep} treated bones as noise, using a combination of autoencoder and deep CNN features to remove bone structures, though this led to image blurring. Inspired by Generative Adversarial Networks (GANs) \cite{goodfellow2014generative}, Zhou \textit{et al.} \cite{zhou2019generation} introduced a Multi-scale Conditional Adversarial Network (MCA-Net) aimed at producing soft tissue images while preserving key anatomical structures. To address the issue of inconsistent background intensity in gradient-based methods, Chen \textit{et al.} \cite{chen2019bone} employed a cascaded CNN into the wavelet domain. Rajaraman \textit{et al.} \cite{rajaraman2021chest} developed the ResNet-BS model for bone suppression in CXR images, validating its effectiveness through subsequent analytic tasks. Liu \textit{et al.} \cite{liu2023bone} proposed a bone suppression technique for lateral CXR images utilizing data rectification and distillation learning. Rani \textit{et al.} \cite{rani2022multi} utilized a Pix2Pix-based conditional Generative Adversarial Network (cGAN) for effective bone suppression. Additionally, Singh \textit{et al.} \cite{singh2024mda} advanced this work with the Multiscale and Dual Attention Generative Adversarial Network (MDA-GAN), which incorporates a multiscale and dual attention mechanism into a conditional generative adversarial network, effectively removing bone structures while preserving background and chest X-ray details.
Furthermore, Schiller \textit{et al.} \cite{schiller2025xu} provided the xU-NetFullSharp architecture with multi-level and multi-directional connections to remove bone shadows in CXR images.

Diffusion models \cite{ho2020denoising} have emerged as a novel class of generative models, adept at producing high-quality images. They effectively address challenges such as mode collapse and convergence issues that are often encountered with GANs. While diffusion models have shown impressive generative abilities in various image generation fields, they encounter specific challenges when applied to bone suppression:

\begin{itemize}
  \item Diffusion models, compared to other generative models, are more computationally intensive due to their larger number of parameters and iterative processing steps. This high demand limits their application in high-resolution imaging. Consequently, in the context of bone suppression, diffusion models often produce images with lower resolution and less detail than needed for clinical applications \cite{weber2023cascaded}.
  \item The training of diffusion models for denoising varies across different frequency components. Specifically, denoising mid- to high-frequency components is generally easier than addressing low-frequency components during the reverse process. This imbalance, coupled with the instability inherent in multi-step sampling, hampers the generation of low-frequency information. This often results in deviations in pixel intensities, affecting brightness and contrast, which can complicate the interpretation of the generated medical images \cite{alnaggar2024efficient}.

\end{itemize}

To apply diffusion models to bone suppression, Chen \textit{et al.} \cite{chen2024bs} introduced BS-Diff, which involves developing a conditional diffusion model in pixel space, enhanced with an Auto-Encoder as an enhancement module. The cascade architecture used in BS-Diff helps mitigate some low-frequency discrepancies typical of diffusion models. However, the model still faces challenges in achieving adequate resolution and fine detail, alongside high computational demands. Moreover, utilizing the cascade architecture at the same resolution may lead to partial information loss while incurring additional computational costs. Consequently, the challenges associated with diffusion models in bone suppression remain only partially addressed.

\begin{figure*}[!tbp]
\centering
\includegraphics[width=0.92\textwidth]{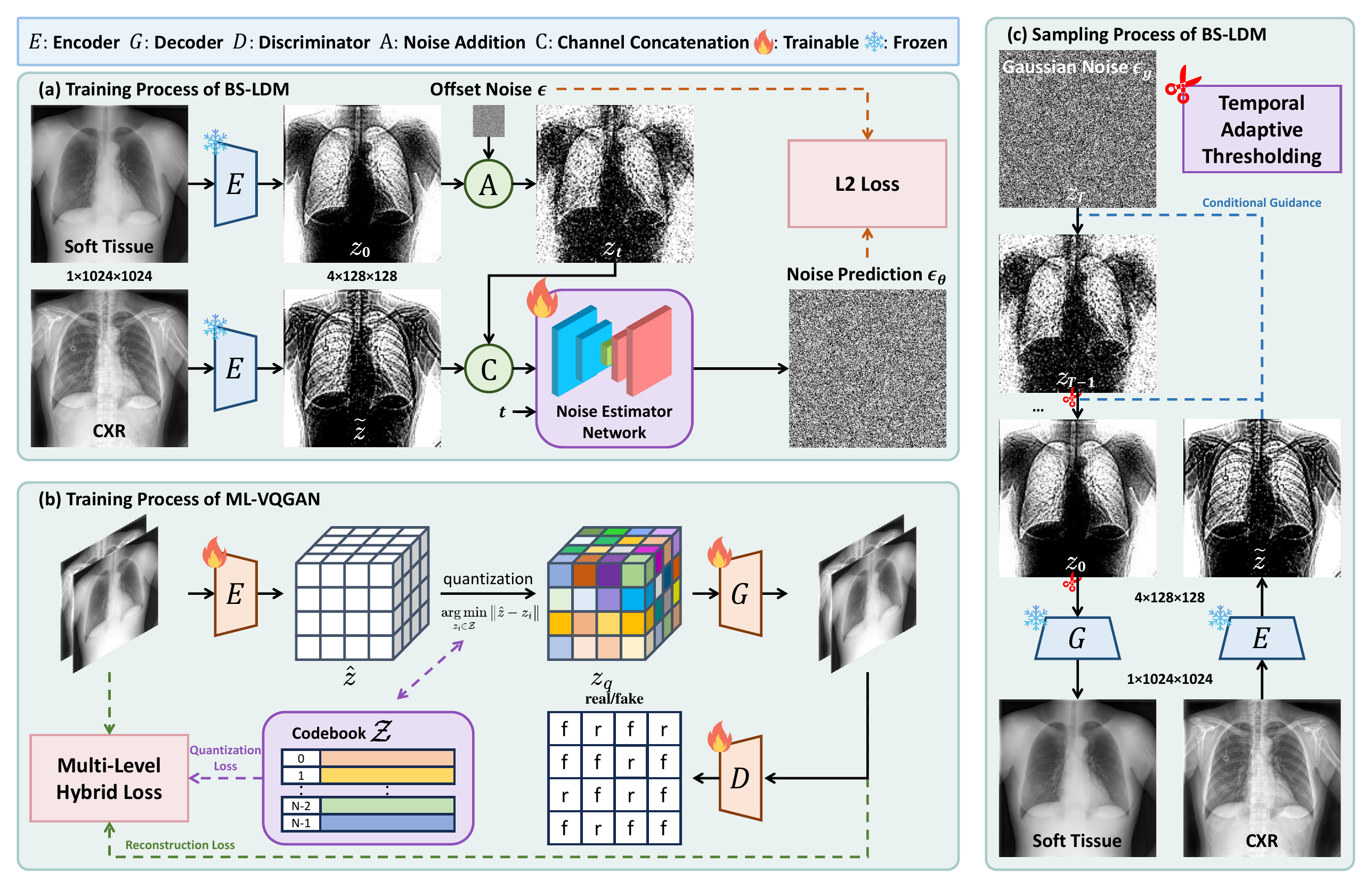}
\caption{Overview of the proposed BS-LDM: (a) The training process of BS-LDM, where CXR and noised soft tissue data in the latent space are transmitted to the noise estimator for offset noise prediction and L2 loss calculation; (b) The training process of ML-VQGAN, where a multi-level hybrid loss function is used to construct a latent space by reconstruction training while preserving texture details; (c) The sampling process of BS-LDM, where the latent variables obtained after each denoising step are clipped using a temporal adaptive thresholding strategy for the sake of contrast stability.}
\label{frame}

\end{figure*}

Recently, Latent Diffusion Models (LDMs) \cite{rombach2022high} have emerged as an extension of diffusion models, operating in a lower-dimensional space to enable the synthesis of high-resolution images with reduced computational costs. By leveraging latent space, these models significantly decrease the number of parameters in the noise estimator network. Building on this innovation, we propose an end-to-end LDM-based framework for high-resolution bone suppression, called BS-LDM, which treats the task as CXR-guided image denoising to achieve effective bone suppression. To enhance detail preservation, we design a Multi-Level hybrid loss-constrained Vector-Quantized Generative Adversarial Network (ML-VQGAN), for perceptual compression of high-resolution images. Additionally, to address discrepancies in low-frequency information generation by compensating low-frequency disruption in the forward process, we introduce offset noise \cite{offset} into the noise addition process to correct luminance offset during image generation. Additionally, we propose a temporal adaptive thresholding strategy to resolve pixel saturation issues during the reverse process. To assess the clinical relevance of our framework in aiding lung disease diagnosis, we conducted a clinical evaluation focusing on image quality and diagnostic utility. The results demonstrated excellent image quality scores and significant diagnostic improvements, highlighting the clinical value of our approach.
Our automated downstream evaluation results further demonstrate the potential of our BS-LDM model for enhancing the performance of computer-assisted lung lesion detection. Furthermore, we compiled a high-quality bone suppression dataset named SZCH-X-Rays, featuring high-resolution paired CXR and DES soft tissue images from 818 patients, collected in collaboration with our partner hospital. We also processed 241 pairs of CXR and DES soft tissue images from the JSRT dataset, the largest open-source dataset for bone suppression, by performing operations like inversion and contrast adjustment to convert these images into negative formats more frequently used in clinical settings. Our primary contributions are summarized as follows:

\begin{itemize}
    
    \item We propose BS-LDM, an LDM-based framework for high-resolution bone suppression in CXR images, and introduce ML-VQGAN for effective perceptual compression and detail retention. To the best of our knowledge, this is the first study to train and evaluate LDMs for the bone suppression task.
    
    \item To enhance the quality of generated images, we incorporate offset noise in the forward process, and temporal adaptive thresholding in the reverse process. These strategies help minimize discrepancies in low-frequency information, thereby improving the interpretability of the soft tissue images.

    \item Our comprehensive experiments, as well as clinical and automated downstream evaluations demonstrated excellent image quality and substantial diagnostic improvements, underscoring the clinical significance of our work.
    
\end{itemize}

\section{Methodology}
\label{sec:methodology}

\subsection{Overall Architecture}
This section presents the proposed BS-LDM framework in detail. It is an end-to-end framework designed for effective bone suppression in high-resolution CXR images using a conditional LDM, as depicted in Fig. \ref{frame}. In this framework, we developed ML-VQGAN for perceptual compression to ensure good detail preservation. Additionally, we incorporated offset noise in the forward process and temporal adaptive thresholding in the reverse process to address discrepancies in low-frequency information generation, enhancing the overall quality of generation. A detailed description of each component of BS-LDM is provided in the following subsections.

\subsection{Conditional Latent Diffusion Models}
LDM-based approaches \cite{blattmann2023align,li2024controlnet++,fang2025lpuwf}, typically operate through a two-stage generative process. First, they map data to a low-dimensional latent space, and then sample from this space to generate the final output. Unlike traditional diffusion models \cite{ho2020denoising}, LDMs significantly reduce computational demands for high-resolution image synthesis by encoding data in a compressed latent space. Specifically, for an image \(x \in \mathbb{R}^{C \times H \times W}\), the encoder \(E\) transforms \(x\) into a latent variable \(z = E(x)\). In the context of conditional LDMs, a forward process, denoted as \(q_\theta(z_t|z_0)\), is defined as a process gradually adding Gaussian noise to the input data \(z_0\): 

\begin{equation}
z_t=\sqrt{\bar{\alpha}_t} z_0+\sqrt{1-\bar{\alpha}_t} \epsilon_g, \quad \epsilon_g \sim \mathcal{N}(\mathbf{0}, \mathbf{I}),
\label{eq_noise}
\end{equation}
where $\epsilon_g$ is a noise map sampled from a Gaussian distribution, and $\bar{\alpha}_t:=\prod_{s=0}^t \alpha_s$. Here, $\alpha_t=1-\beta_t$ is a differentiable function of timestep $t$ determined by the denoising sampler. The diffusion training loss with a given image condition $\widetilde{z}$ at timestep $t$ is formulated as:

\begin{equation}
\mathcal{L}\left(\epsilon_\theta\right)=\mathbb{E}_{z_0, t, \widetilde{z}, \epsilon_g}\left[\left\|\epsilon_\theta\left(z_t, t, \widetilde{z}\right)-\epsilon_g\right\|_2^2\right].
\label{eq_c_loss}
\end{equation}

During the reverse process, starting with random noise $z_T \sim \mathcal{N}(\mathbf{0}, \mathbf{I})$, the final denoised image $x_0$ is predicted through a step-by-step denoising process:
\begin{equation}
z_{t-1}=\frac{1}{\sqrt{\alpha_t}}\left(z_t-\frac{1-\alpha_t}{\sqrt{1-\bar{\alpha}_t}} \epsilon_\theta\left(z_t, t\right)\right)+\sigma_t \epsilon_g,
\label{eq_inf}
\end{equation}
where $\epsilon_\theta$ represents the predicted noise by the noise estimator network with parameters $\theta$ at timestep $t$, and $\sigma_t=\frac{1-\bar{\alpha}_{t-1}}{1-\bar{\alpha}_t} \beta_t$ is the variance of posterior Gaussian distribution $p_\theta\left(x_0\right)$.

In this work, we employ a U-Net with multi-resolution attention as the noise estimator network. The conditional guidance of CXR images in BS-LDM is achieved through channel concatenation of \(z_t\) and \(\widetilde{z}\), as illustrated in Fig. \ref{frame} (a).

\subsection{Multi-Level Hybrid Loss-Constrained VQGAN}
To construct a low-dimensional latent space for LDMs while ensuring high-quality perceptual compression and reconstruction, we introduce a multi-level hybrid loss-constrained VQGAN, named ML-VQGAN, as shown in Fig. \ref{frame} (b). This approach compresses the pixel space \(X \in \mathbb{R}^{1 \times H \times W}\) into a low-dimensional latent space \(Z \in \mathbb{R}^{C \times \frac{H}{r} \times \frac{W}{r}}\) using ML-VQGAN. To prevent the creation of latent spaces with excessively high variance, VQ regularization is applied, allowing for the fine-tuning of latent features and achieving the final coding vectors \(z_q\). 

The total loss function for ML-VQGAN is defined as follows:
\begin{equation}
\mathcal{L}_{Total} = \mathcal{L}_{Recon} + \lambda_{Qua} \cdot \mathcal{L}_{Qua},
\label{eq_tloss}
\end{equation}
where
\begin{equation}
\mathcal{L}_{Recon} = \lambda_{L1} \cdot \mathcal{L}_{L1} + \lambda_{Per} \cdot \mathcal{L}_{Per} + \lambda_{Adv} \cdot \mathcal{L}_{Adv}.
\label{eq_rloss}
\end{equation}
In this formulation, \(\mathcal{L}_{Recon}\) is a combination of different reconstruction losses at multiple levels. The weight coefficients \(\lambda_{Qua}\), \(\lambda_{L1}\), \(\lambda_{Per}\), and \(\lambda_{Adv}\) adjust the influence of the respective components \(\mathcal{L}_{Qua}\), \(\mathcal{L}_{L1}\), \(\mathcal{L}_{Per}\), and \(\mathcal{L}_{Adv}\) on the total loss.

\subsubsection{Reconstruction Loss}
To achieve high-quality image reconstruction, we incorporate L1 loss \(\mathcal{L}_{L1}\), perceptual loss \(\mathcal{L}_{Per}\) \cite{johnson2016perceptual}, and adversarial loss \(\mathcal{L}_{Adv}\) \cite{wang2018pix2pixHD} into the reconstruction loss \(\mathcal{L}_{Recon}\). The L1 loss \(\mathcal{L}_{L1}\) evaluates the pixel-level differences between the reconstructed image and the original image, defined as:
\begin{equation}
\mathcal{L}_{L1}=\frac{1}{N} \sum_{i=1}^N|x_i-\hat{x}_i|.
\label{eq_l1}
\end{equation}
To address the potential blurring effects of relying solely on pixel-level loss, we introduce the perceptual loss \(\mathcal{L}_{Per}\), which uses features from pre-trained deep neural networks to enhance image quality:
\begin{equation}
\mathcal{L}_{Per}=\frac{1}{N} \sum_{k=1}^N\left(\phi_k(x)-\phi_k(\hat{x})\right)^2,
\label{eq_per}
\end{equation}
where \(\phi_k\) denotes the \(k\)-th feature map extracted from a pre-trained deep neural network, which, in our study, is the VGG16 model \cite{simonyan2014very}.

Additionally, to produce more realistic images, we include the adversarial loss \(\mathcal{L}_{Adv}\), which employs a discriminator network \(D\) to differentiate between real and generated images:
\begin{equation}
\mathcal{L}_{Adv}=-\mathbb{E}[\log D(G(x))],
\label{eq_adv}
\end{equation}
where $D(G(x))$ represents the probability that generated data is perceived as real.
\subsubsection{Quantization Loss}
In ML-VQGAN, the concept of quantization loss, \(\mathcal{L}_{Qua}\), is used to optimize the codebook \(\mathcal{Z}\). The mathematical expression for this loss is:

\begin{equation}
\mathcal{L}_{Qua} = \left\|sg(E(x))-z_q\right\|_2^2+\beta\left\|E(x)-sg\left(z_q\right)\right\|_2^2,
\label{eq_qua}
\end{equation}
where $sg$ signifies the stop-gradient operation, and \(\beta\) represents the weight used to balance the optimization between the encoder \(E\) and the codebook \(\mathcal{Z}\).

\subsection{Offset Noise}

\begin{figure}[!tbp]
\centering
\includegraphics[width=0.9\columnwidth]{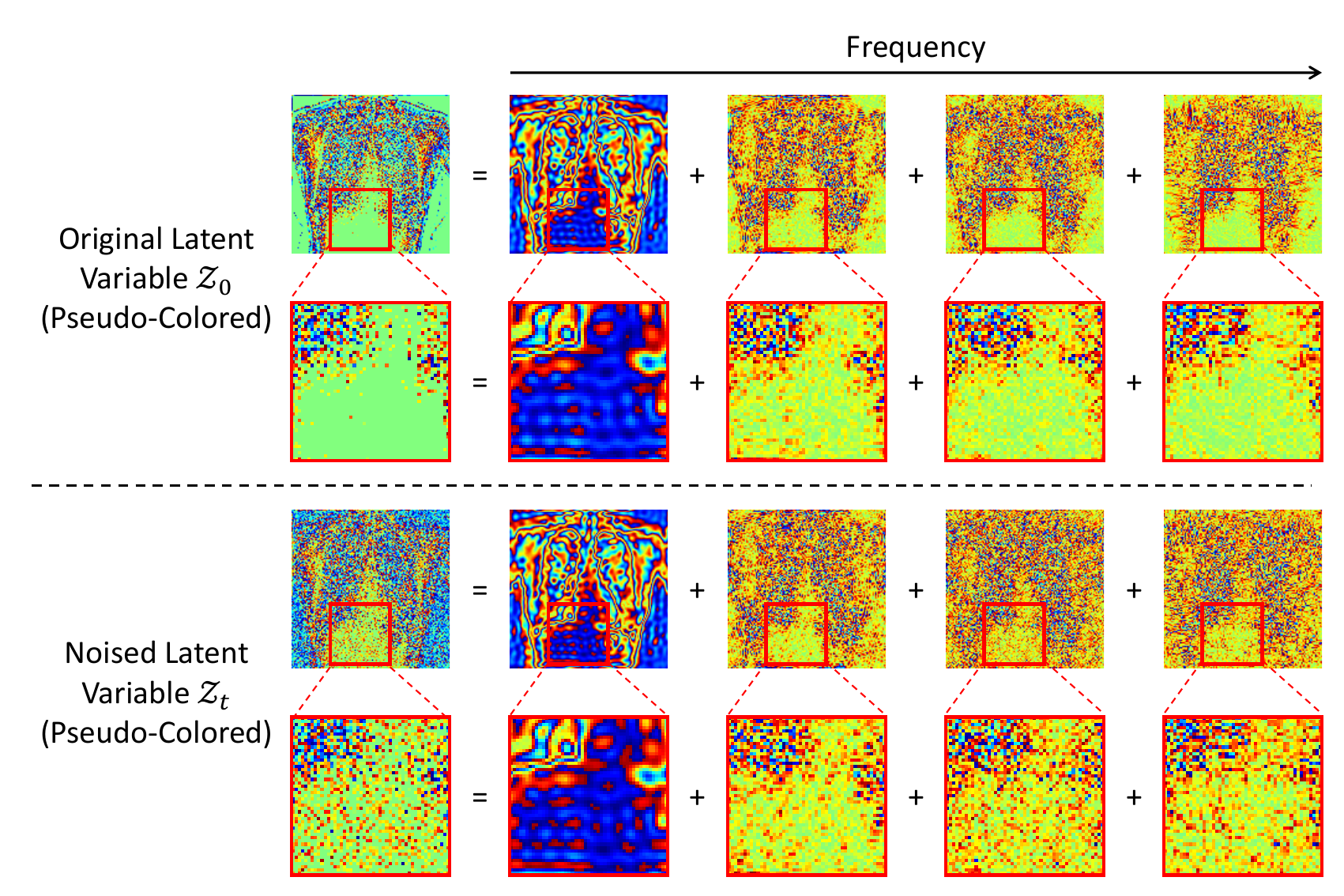}
\caption{Visualization of high-frequency and low-frequency feature decomposition of latent variables before and after Gaussian noise addition using Discrete Fourier Transform. The results are pseudo-colored for ease of demonstration.} \label{freq}
\vspace{-0.5cm}

\end{figure}

\begin{figure}[!tbp]
\centering
\includegraphics[width=0.9\columnwidth]{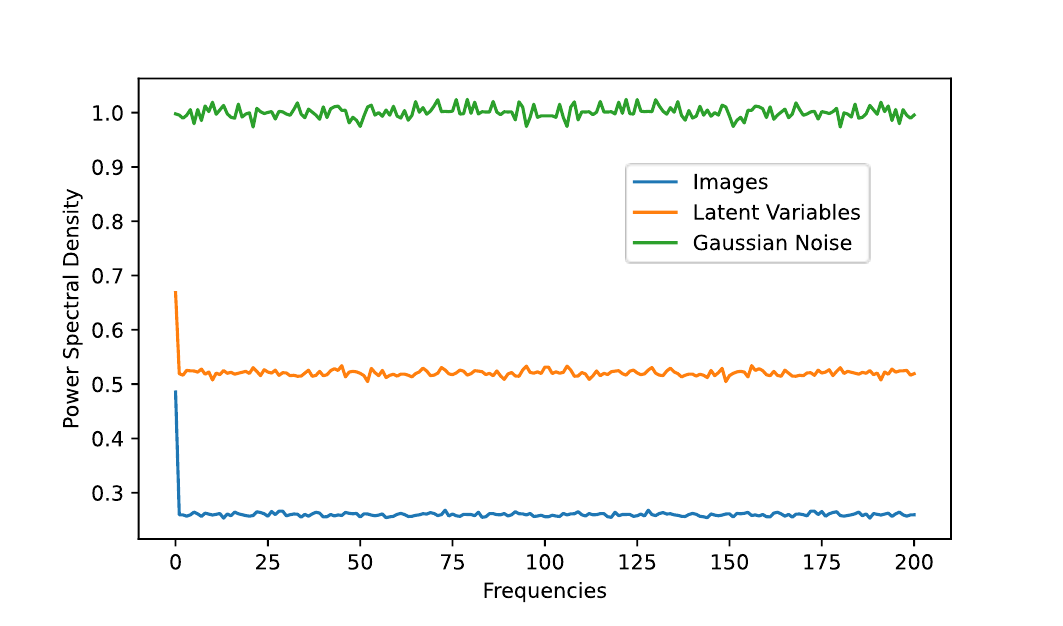}
\caption{Power spectral densities of soft tissue images in SZCH-X-Rays, corresponding latent variables and Gaussian noise on 201 spectrogram components, averaged over 10000 samples.} \label{psd}
\vspace{-0.5cm}

\end{figure}

In the forward process of LDMs, each pixel gradually accumulates a small amount of independently and identically distributed (i.i.d.) Gaussian noise at each step. LDMs are trained to predict this noise after it has disrupted real images. 
Nervertheless, when implementing bone suppression using LDMs, we observed a notable discrepancy in their performance: while LDMs effectively generated high-frequency textures, their handling of low-frequency information was suboptimal. To investigate this issue, we conducted a spectral decomposition analysis of the latent variables of soft tissue images using the Discrete Fourier Transform. This analysis revealed differences in how conventional Gaussian noise injection impacts frequency bands during the generative process. As illustrated in Fig.~\ref{freq}, low-frequency features exhibit greater resistance to Gaussian noise compared to high-frequency features. This disparity affects the training dynamics of the noise estimator network, leading to uneven estimation capabilities across frequency components. Specifically, the reduced ability to denoise low-frequency features forces LDMs to retain certain elements of the initialization point, namely, Gaussian noise, as low-frequency information during the reverse generation process. Consequently, this manifests as a shift in the mean value of the generated images towards the mean of the Gaussian noise. To further explore the underlying causes of this phenomenon, we calculated the power spectral densities of soft tissue images from the SZCH-X-Rays dataset, alongside their corresponding latent variables and Gaussian noise across various frequencies. The results, presented in Fig.~\ref{psd}, provide deeper insights into the interplay between frequency components and Gaussian noise during the LDM generative process. The results demonstrate that the power spectral density of Gaussian noise is relatively uniform across all frequencies, with slight fluctuations around a constant value. 
This indicates that the Gaussian noise has a consistent intensity at different frequencies. In contrast, the power spectral density of soft tissue images and their corresponding latent variables exhibits a distinct pattern. 
The analysis showed that the power spectral density is higher in the low-frequency region and decreases progressively in the mid- to high-frequency regions. This indicates a relatively greater intensity of low-frequency information compared to higher-frequency components.
This difference can be attributed to the inherent characteristics of soft tissue images and the nature of Gaussian noise. Soft tissue images contain rich low-frequency information which represent the overall shape and intensity variations of the tissues. These low-frequency features are crucial for the visual perception and diagnostic value of the images.

\begin{figure}[!tbp]
\centering
\includegraphics[width=0.8\columnwidth]{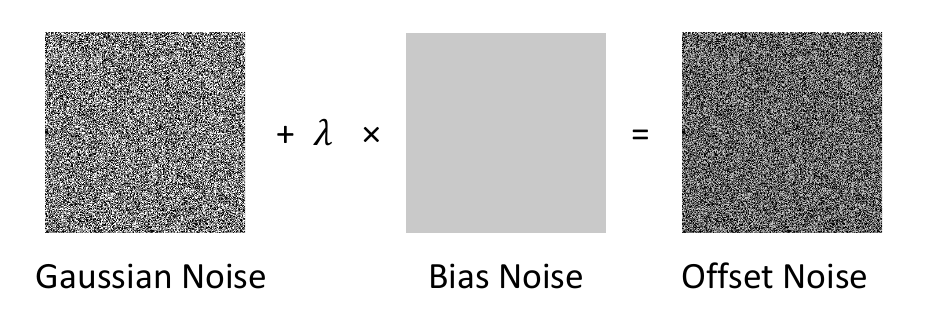}
\caption{Illustration of the composition of offset noise.} \label{off}
\vspace{-0.5cm}

\end{figure}

To address this issue, we apply offset noise \(\epsilon\) into the forward process to compensate the disturbance of low-frequency information, as illustrated in Fig.~\ref{frame}(a). This approach involves incorporating additional zero-frequency bias noise into the standard noise injection procedure, as demonstrated in Fig.~\ref{off} and mathematically defined in Equation (\ref{eq9}):
\begin{equation}
z_t=\sqrt{\bar{\alpha}_t} z_0+\sqrt{1-\bar{\alpha}_t} \epsilon, \quad \epsilon \sim \mathcal{N}(0, \mathbf{I} + \lambda \cdot \Sigma),
\label{eq9}
\end{equation}
where \(\Sigma\) is a covariance matrix of all ones, and \(\lambda\) is the weight of the bias noise. Bias noise is essentially zero-frequency noise, which can be derived from a one-pixel Gaussian noise using tensor broadcasting. The introduction of offset noise aims to completely eliminate the low-frequency information from real images during the noise addition process and ensure that the endpoints converge to true Gaussian noise. This modification broadens the spectrum of mean value variations in the generated images, thereby enhancing their resemblance to real images.

\subsection{Temporal Adaptive Thresholding Strategy}
In the reverse diffusion process, it is standard practice to constrain predicted values within the range of \([-1, 1]\) to maintain numerical stability. This approach, referenced in prior studies \cite{ho2020denoising}, has not received extensive attention. Saharia \textit{et al.} referred to this as static thresholding and introduced a dynamic thresholding method to tackle the pixel saturation issues common in classifier-free guidance \cite{cfg}. They employed a strategy where a threshold, \(s\), was set to a particular percentile of the absolute pixel values, clipping data to \([-s, s]\) and normalizing by \(s\) if \(s > 1\).

Although thresholding is typically absent in latent diffusion models \cite{lu2022dpm}, we observed that without it, these models are prone to pixel saturation, resulting in overly contrasted outputs. While static thresholding provides numerical stability and helps mitigate pixel saturation, it significantly limits image variance in our task. This limitation results in noticeable contrast differences between the generated images and real ones, as shown in Fig.~\ref{abl_img}. Similarly, while dynamic thresholding provides more flexibility, it still fails to adequately generate soft tissue images with clear interpretability. In LDMs, the distribution of the latent variable \(z_t\) changes over time as \(t\) progresses. Applying static or dynamic thresholding based solely on pixel statistics can introduce biased contrast, as it fails to account for the temporal dynamics of the latent variable.

To address these issues and align pixel intensities of generated soft tissue images with real ones, we propose a temporal adaptive thresholding strategy. This method dynamically adjusts the thresholding range \([-s, s]\) according to the current sampling step \(t\). For the thresholding operation $\tau$, we utilize a linear function to define the relationship between the threshold \(s\) and current sampling timestep \(t\), as shown in Equation (\ref{eq_s}):
\begin{equation}
\tau(z_{t,i}, s) = 
\begin{cases} 
-s, & \text{if } z_{t,i} < -s \\
z_{t,i}, & \text{if } -s \leq z_{t,i} \leq s \\
s, & \text{if } z_{t,i} > s
\end{cases}, \quad s = \omega \cdot t + b,
\label{eq_s}
\end{equation}
where \(z_{t,i}\) represents the \(i\)-th pixel  of latent variable \(z_t\) at timestep \(t\), \(\omega (> 0)\) is the slope, and \(b (\geq 1)\) is the intercept of the linear function.

By progressively expanding the thresholding range as the sampling evolves, temporal adaptive thresholding affords the model enhanced flexibility. This enables a nuanced adjustment of pixel intensities, ensuring the contrast of the final generated soft tissue images closely resembles that of real images.

\section{Experiments and Discussion}
\subsection{Data Preparation and Preprocessing}
We compiled a dataset of 831 paired posterior-anterior CXR and DES soft tissue images. These images were acquired using a digital radiography system with a dual-exposure DES unit (Discovery XR656, GE Healthcare) from our partner hospital. Originally archived in 14-bit DICOM format, the images were converted to PNG format to simplify processing procedures. Each image measures \(2021 \times 2021\) pixels, with pixel sizes ranging from 0 to 0.1943 mm. From this collection, we excluded 13 image pairs due to operational errors, severe motion artifacts, and conditions like pleural effusions and pneumothorax, which could interfere with analysis. The refined dataset, named SZCH-X-Rays, consists of 818 image pairs. These were divided into training, validation, and test sets in an 8:1:1 ratio. We enhanced local contrast in the images using contrast-limited adaptive histogram equalization.

Additionally, we processed 241 pairs of CXR and DES images from the JSRT dataset, the largest open-source collection available. Operations included inversion and contrast adjustment to convert them into negative images, commonly used in clinical practice. To optimize memory usage, images from both SZCH-X-Rays and JSRT datasets were resized to \(1024 \times 1024\) pixels. Finally, all pixel values were normalized to the range \([-1, 1]\).

\subsection{Implementation Details}
All experiments were carried out using PyTorch 2.0.1 on a single Nvidia A100 80G GPU within Ubuntu 20.04 systems. We trained the ML-VQGAN model from scratch for 1,000 epochs with a batch size of 4, employing the Adam optimizer. Similarly, we trained our BS-LDM model from scratch for 2,500 epochs with a batch size of 4 using the AdamW optimizer, which required 22 hours.

For ML-VQGAN training, the loss weights were set to \(\lambda_{L1} = \lambda_{Qua} = 1\), \(\lambda_{Per} = 0.001\), and \(\lambda_{Adv} = 0.01\), completing the process in approximately 20 hours. During BS-LDM training, we implemented an Exponential Moving Average (EMA) strategy with a decay coefficient of 0.995. We also employed a dynamic learning rate schedule, starting at 0.0001 for the ML-VQGAN encoder \(E\) and decoder \(G\), 0.0005 for the ML-VQGAN discriminator \(D\), and 0.0002 for the BS-LDM noise estimator network.

For BS-LDM, the number of training and inference timesteps \(T\) was set to 1,000. The variance schedule parameter \(\beta\) varied from 0.008 to 0.02, distributed across \(T\) timesteps using a cosine noise schedule. In the forward process, the weight of bias noise in offset noise, \(\lambda\), was set at 0.1. In the reverse process, we adopted a time-adaptive thresholding strategy, selecting \(\omega = 0.003\) and \(b = 1.4\) for the linear function's slope and intercept, respectively.

\begin{figure*}[!tbp]
\centering
\includegraphics[width=0.9\textwidth]{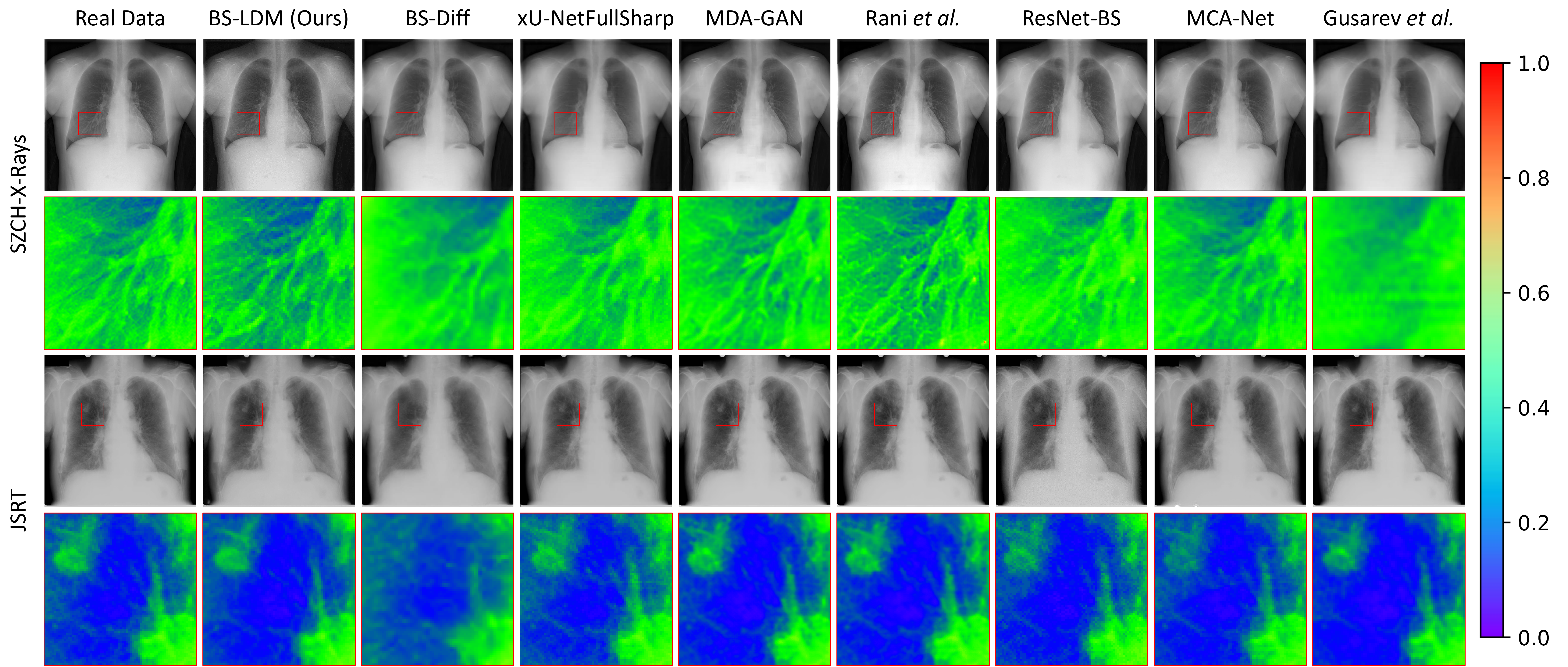}
\caption{Bone suppression and pseudo-color zoomed-in views of BS-LDM compared with state-of-the-art approaches on SZCH-X-Rays (top) and JSRT (bottom) datasets.} \label{comp}
\vspace{-0.3cm}
\end{figure*}

\subsection{Evaluation Metrics}
We evaluate the quality of bone suppression using four metrics: Bone Suppression Ratio (BSR) \cite{BSR}, Mean Squared Error (MSE), Peak Signal-to-Noise Ratio (PSNR) \cite{hore2010image}, and Learned Perceptual Image Patch Similarity (LPIPS) \cite{zhang2018unreasonable}. These metrics are applied in both comparative analyses and ablation studies.

The BSR evaluates how effectively bone suppression is applied in CXR images. It measures the difference between the estimated soft tissue image, denoted as $\hat{S}$, and the actual soft tissue image, $S$, while considering the corresponding bone image $B$. The BSR formula is expressed as follows:

\begin{equation}
BSR=1-\sum_{i=1}^N(S_i-\hat{S}_i)^2 /  \sum_{i=1}^N B_i^2,
\label{eq_bsr}
\end{equation}
where a BSR value of 1 indicates ideal performance. In addition to BSR, we employ two common pixel-level metrics, MSE and PSNR, which assess the pixel-level discrepancies between the generated and actual soft tissue images. To further evaluate the generation of detail from a perceptual standpoint, we utilize LPIPS. This metric measures perceptual similarity between two images by comparing features extracted from a pre-trained deep neural network. LPIPS is calculated as follows:
\begin{equation}
LPIPS=\frac{1}{N\cdot M\cdot F} \sum_{i=1}^N \sum_{j=1}^M \sum_{f=1}^F
\left(\phi_{i, j, f}(S)-\phi_{i, j, f}(\hat{S})\right)^2,
\label{eq_lpips}
\end{equation}
where $N \cdot M$ represents the number of image patches, $F$ denotes the number of feature channels in the selected network layer, and $\phi_{i, j, f}$ indicates the features at position $(i,j)$ and channel $f$.

To evaluate the downstream diagnostic value of our bone suppression method, we employ five metrics: Precision, Sensitivity, F1 Score, Accuracy, and Specificity. Precision quantifies the proportion of true positives among all positive classifications, minimizing false positives. Sensitivity (Recall) measures the method's ability to detect true pathological findings in bone-suppressed images, which is particularly important for disease screening applications. The F1 Score balances Precision and Sensitivity, providing a comprehensive metric that mitigates potential biases arising from class imbalances. Accuracy reflects the overall correctness of diagnostic predictions across all categories, ensuring clinical reliability. Lastly, Specificity evaluates the method's ability to correctly identify normal anatomical structures, minimizing false alarms and preserving accuracy in healthy case interpretations.

\subsection{Comparison Results}
To ensure a fair and comprehensive evaluation, we compared our proposed model against several state-of-the-art approaches recently highlighted in the literature. Specifically, our assessment included comparisons with an autoencoder-based model \cite{gusarev2017deep}, the GAN-based MDA-GAN \cite{singh2024mda}, method by Rani \textit{et al.} \cite{rani2022multi} and MCA-Net \cite{zhou2019generation}, the ResNet-based ResNet-BS \cite{rajaraman2021chest}, the U-Net-based xU-NetFullSharp \cite{schiller2025xu}, and the diffusion model-based BS-Diff \cite{chen2024bs}.

For consistency, we used the default parameters from the open-source implementations of these models and maintained the same resolution across all experiments. The comparative performance of these bone suppression techniques is shown for the SZCH-X-Rays and JSRT datasets in Table \ref{compt} and visually represented in Fig. \ref{comp}.

\begin{figure}[!tbp]
\centering
\includegraphics[width=0.9\columnwidth]{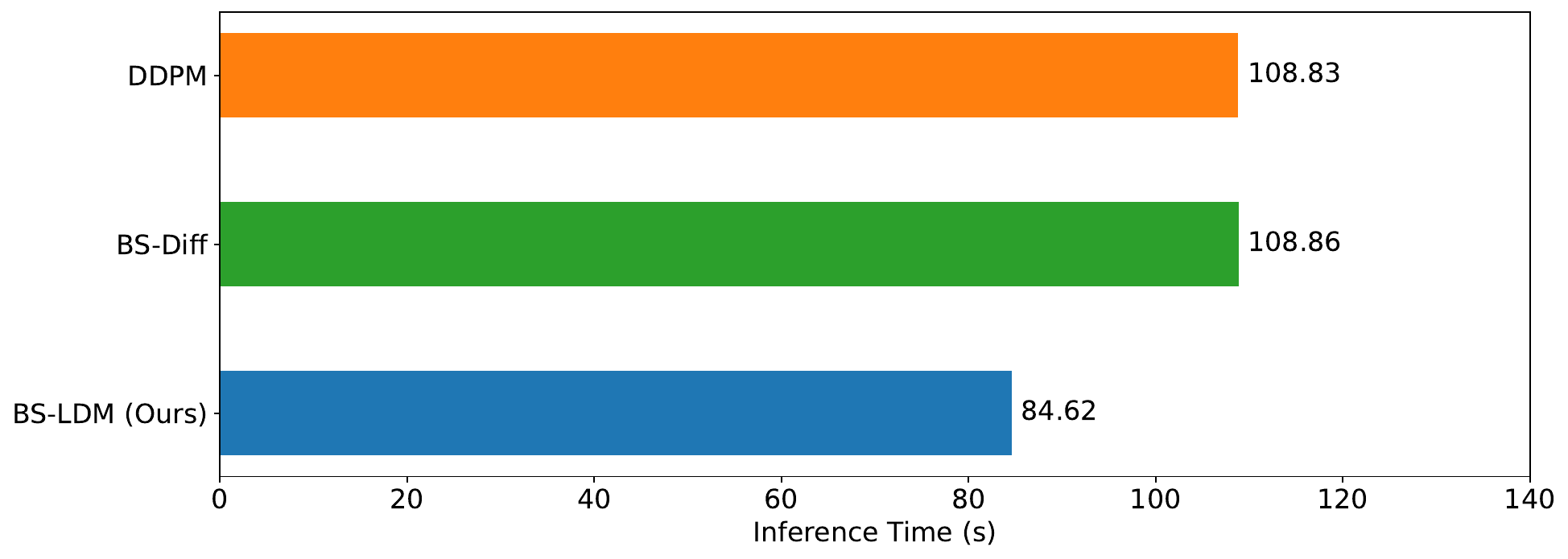}
\caption{Inference efficiency comparison of our method and other diffusion-based methods on the SZCH-X-Rays dataset on a single Nvidia A100 80G GPU.} \label{bar}
\vspace{-0.5cm}
\end{figure}

\subsubsection{Evaluation on the SZCH-X-Rays Dataset}
In our SZCH-X-Rays dataset, the proposed method demonstrates superior performance across all evaluation metrics compared to existing approaches. Specifically, it achieves improvements in BSR by 1.6\% ($\uparrow$), MSE by 16.7\% ($\downarrow$), PSNR by 1.037 dB ($\uparrow$), and LPIPS by 34.6\% ($\downarrow$). The reduction in LPIPS, exceeding 34.6\%, underscores the enhanced retention of critical lung details, as illustrated in Fig. \ref{comp}. Among the evaluated methods, the autoencoder-based model and BS-Diff perform poorly across most metrics, largely due to significant texture loss. While MDA-GAN and xU-NetFullSharp exhibit relatively better detail retention, they fall short of the proposed BS-LDM method due to minor inconsistencies in brightness and detail preservation. Notably, all methods achieve effective bone suppression, with a minimum BSR of 94.4\%. However, BS-LDM surpasses them with a BSR of 97.6\%. Additionally, as shown in Fig. \ref{bar}, BS-LDM demonstrates efficient processing with an inference time approximately 77.7\% of that required by BS-Diff and DDPM, further solidifying its practicality for clinical applications.

\subsubsection{Evaluation on the JSRT Dataset}
On the public JSRT dataset, our method consistently outperforms all other approaches across key metrics, achieving improvements in BSR by 2.1\% ($\uparrow$), MSE by 11.3\% ($\downarrow$), PSNR by 2.315 dB ($\uparrow$), and LPIPS by 31.0\% ($\downarrow$). As illustrated in Fig. \ref{comp}, BS-LDM produces sharper and more detailed outputs compared to BS-Diff and MCA-Net. However, the distinctions are less pronounced than those observed in the SZCH-X-Rays dataset, likely due to the inherent blurring present in the JSRT dataset. Quantitatively, BS-LDM delivers significantly lower LPIPS scores on the JSRT dataset, reflecting enhanced perceptual quality and better preservation of soft tissue details. By contrast, the autoencoder-based model, MCA-Net, Rani \textit{et al.}, and BS-Diff demonstrate mediocre performance across most metrics, which can be attributed to minor pixel-level luminance inconsistencies and a lack of fine detail retention.

\begin{figure*}[!thbp]
\centering
\includegraphics[width=0.9\textwidth]{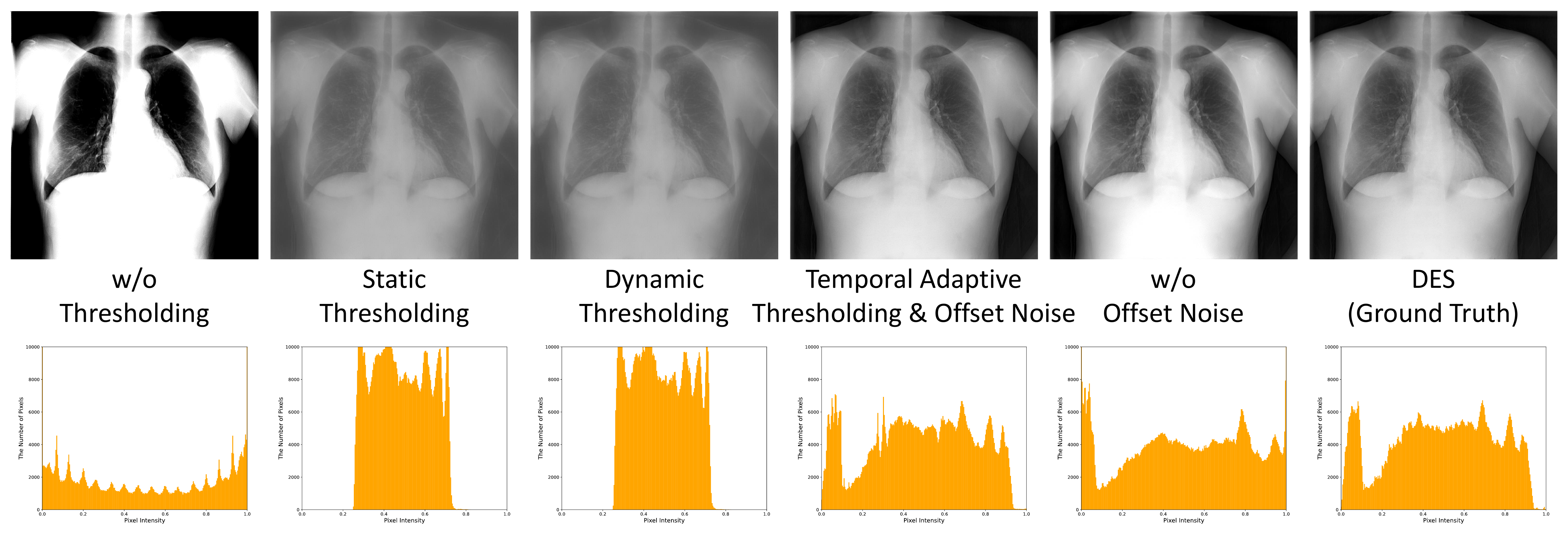}
\caption{Visualization of ablation studies on BS-LDM. The upper row shows generated soft tissue images under different conditions, while the lower row presents corresponding pixel intensity histograms, where rightward shift indicates brighter images, and wider spread indicates higher contrast.}
\label{abl_img}
\end{figure*}

\begin{table*}[!thbp]
\caption{Comparison on SZCH-X-Rays and JSRT datasets with the state-of-the-art methods using four evaluation metrics.}\label{compt}
\centering
\begin{tabular}{p{3cm}<{\centering}p{3cm}<{\centering}p{2.3cm}<{\centering}p{2.3cm}<{\centering}p{2.3cm}<{\centering}p{2.3cm}<{\centering}}
\toprule
Dataset & Method &  BSR ($\uparrow$) & MSE ($\downarrow$) & PSNR ($\uparrow$) & LPIPS ($\downarrow$)\\
\midrule
\multirow{8}{*}{SZCH-X-Rays}& Gusarev \textit{et al.} \cite{gusarev2017deep}  &  0.944 $\pm$ 0.018 & 0.00084 $\pm$ 0.00023 & 31.900 $\pm$ 3.449 & 0.220 $\pm$ 0.012\\
&MCA-Net \cite{zhou2019generation} &  0.945 $\pm$ 0.019 & 0.00073  $\pm$ 0.00021& 32.177 $\pm$ 2.912  & 0.118 $\pm$ 0.011\\
&ResNet-BS \cite{rajaraman2021chest} &  0.956 $\pm$ 0.021 & 0.00085 $\pm$ 0.00034 & 31.850 $\pm$ 3.097 & \underline{0.078 $\pm$ 0.008}\\
&Rani \textit{et al.} \cite{rani2022multi} &  0.955 $\pm$ 0.031 & 0.00082 $\pm$ 0.00039 & 31.961 $\pm$ 2.924 & 0.084 $\pm$ 0.022\\
&MDA-GAN \cite{singh2024mda} &  0.957 $\pm$ 0.025 & 0.00075 $\pm$ 0.00026 & 32.180 $\pm$ 3.077& 0.128 $\pm$ 0.026\\
&xU-NetFullSharp \cite{schiller2025xu} &  0.950 $\pm$ 0.014 & \underline{0.00072 $\pm$ 0.00038} & \underline{32.187 $\pm$ 3.747} & 0.125 $\pm$ 0.024\\
&BS-Diff \cite{chen2024bs} &  \underline{0.961 $\pm$ 0.022} & 0.00077 $\pm$ 0.00043 & 32.181 $\pm$ 3.296 & 0.191 $\pm$ 0.015\\
&\textbf{BS-LDM (Ours)}  &  \textbf{0.976 $\pm$ 0.018} & \textbf{0.00060 $\pm$ 0.00032} & \textbf{33.224 $\pm$ 3.577}  & \textbf{0.051 $\pm$ 0.016}\\
\midrule
\multirow{8}{*}{JSRT}& Gusarev \textit{et al.} \cite{gusarev2017deep} &  0.865 $\pm$ 0.041 & 0.00094 $\pm$ 0.00024& 30.006 $\pm$ 2.234  & 0.193 $\pm$ 0.017\\
&MCA-Net \cite{zhou2019generation} &  0.892 $\pm$ 0.039 & 0.00090 $\pm$ 0.00023 & 31.215 $\pm$ 1.576  & 0.148 $\pm$ 0.017\\
&ResNet-BS \cite{rajaraman2021chest} &  \underline{0.903 $\pm$ 0.038} & 0.00089 $\pm$ 0.00029 & 31.239 $\pm$ 1.639 & 0.139 $\pm$ 0.020\\
&Rani \textit{et al.} \cite{rani2022multi} &  0.896 $\pm$ 0.037 & 0.00087 $\pm$ 0.00031 & 31.435 $\pm$ 2.246 & 0.081 $\pm$ 0.016\\
&MDA-GAN \cite{singh2024mda} &  0.894 $\pm$ 0.035 & 0.00085 $\pm$ 0.00034 & 31.671 $\pm$ 2.277& 0.082 $\pm$ 0.018\\
&xU-NetFullSharp \cite{schiller2025xu} &  0.900 $\pm$ 0.055 & 0.00084 $\pm$ 0.00045 & 31.868 $\pm$ 3.879 & \underline{0.071 $\pm$ 0.017}\\
&BS-Diff \cite{chen2024bs} &  0.902 $\pm$ 0.032 & \underline{0.00080 $\pm$ 0.00033} & \underline{31.997 $\pm$ 2.555}  & 0.147 $\pm$ 0.024\\
&\textbf{BS-LDM (Ours)}  &  \textbf{0.922 $\pm$ 0.028 } & \textbf{0.00071 $\pm$ 0.00024} & \textbf{34.312 $\pm$ 2.176}  & \textbf{0.049 $\pm$ 0.018}\\
\bottomrule
\end{tabular}
\end{table*}

\begin{table*}[!thbp]
\caption{Ablation performance on SZCH-X-Rays and JSRT datasets using three evaluation metrics.}\label{abl}
\centering
\begin{tabular}{p{1.5cm}<{\centering}p{4.5cm}<{\centering}p{1.4cm}<{\centering}p{1.4cm}<{\centering}p{1.4cm}<{\centering}p{1.4cm}<{\centering}p{1.4cm}<{\centering}p{1.4cm}<{\centering}}
\toprule
\multirow{2.4}{*}{Offset Noise} & \multirow{2.4}{*}{Thresholding Strategy} & \multicolumn{3}{c}{SZCH-X-Rays} & \multicolumn{3}{c}{JSRT} \\
\cmidrule(lr){3-8}
&&MSE ($\downarrow$) & PSNR ($\uparrow$) & LPIPS ($\downarrow$) & MSE ($\downarrow$) & PSNR ($\uparrow$) & LPIPS ($\downarrow$) \\
\midrule
$\times$ & $\times$ & 0.1131 & 30.054 & 0.553 & 0.2619 & 28.109 & 0.918\\
$\times$ & Static Thresholding & 0.0164 & 30.222 & 0.141 & 0.0140 & 28.525 & 0.083\\
$\times$ & Dynamic Thresholding & 0.0161 & 30.319 & 0.142 & 0.0123 & 29.368 & 0.081\\
$\times$ & Temporal Adaptive Thresholding & \underline{0.0080} & \underline{31.557} & \underline{0.053} & \underline{0.0109} & \underline{33.415} & \underline{0.053}\\
$\surd$ & $\times$ & 0.1105 & 30.056 & 0.549 & 0.2590 & 28.104 & 0.809\\
$\surd$ & Static Thresholding & 0.0106 & 31.441 & 0.128 & 0.0140 & 28.525 & 0.083\\
$\surd$ & Dynamic Thresholding & 0.0103 & 29.937 & 0.130 & 0.0153 & 29.040 & 0.072 \\
\textbf{$\surd$} & \textbf{Temporal Adaptive Thresholding} & \textbf{0.0006} & \textbf{33.224} & \textbf{0.051} & \textbf{0.0007} & \textbf{34.312}  & \textbf{0.049}\\
\bottomrule
\end{tabular}
\end{table*}

\begin{table}[!thbp]
\caption{Impact of different weight schemes of the multi-level hybrid loss function on SZCH X-Rays dataset.}\label{hyptab1}
\centering
\begin{tabular}{>{\centering\arraybackslash}m{2.8cm} >{\centering\arraybackslash}m{1.4cm} >{\centering\arraybackslash}m{1.4cm} >{\centering\arraybackslash}m{1.4cm}}
\toprule
\multicolumn{1}{c} {Weight Scheme $^\star$}  & MSE ($\downarrow$) & PSNR ($\uparrow$) & LPIPS ($\downarrow$) \\
\midrule
\multirow{1}{2.8cm}{[1, 1, 1, $10^{-2}$]}&0.00073&30.013&0.060
\\
\multirow{1}{2.8cm}{[1, 1, 0, $10^{-2}$]} & 0.00063 & 32.976 & 0.133\\
\multirow{1}{2.8cm}{[1, 1, $10^{-1}$, $10^{-2}$]}& 0.00076	& 29.462 & \underline{0.052}\\
\multirow{1}{2.8cm}{[1, 1, $10^{-2}$, $10^{-2}$]}&0.00062&32.525&0.057
\\
\multirow{1}{2.8cm}{\textbf{[1, 1, \boldmath{$10^{-3}$}, \boldmath{$10^{-2}$}]}} & \textbf{0.00060} & \textbf{33.224} & \textbf{0.051}\\
\multirow{1}{2.8cm}{[1, 1, $10^{-3}$, $10^{-1}$]} &\underline{0.00061}&\underline{33.199}&0.144
\\
\multirow{1}{2.8cm}{[1, 1, $10^{-3}$, 0]} &0.00062&32.845&0.142
\\
\multirow{1}{2.8cm}{[1, 1, $10^{-3}$, 1]} &0.00076&29.673&0.196
\\
\multirow{1}{2.8cm}{[1, 0, $10^{-3}$, $10^{-2}$]} &0.00390&27.130&0.185
\\
\multirow{1}{2.8cm}{[0, 1, $10^{-3}$, $10^{-2}$]} &0.00128&28.743&0.062
\\
\bottomrule
\multicolumn{4}{l}{\small $\star$ $\lambda_{L1}$, $\lambda_{Qua}$, $\lambda_{Per}$ and $\lambda_{Adv}$, from left to right.}\\
\end{tabular}
\label{hyp1}
\vspace{-0.5cm}

\end{table}

\begin{table}[!thbp]
\caption{Impact of different weight schemes of the multi-level hybrid loss function on JSRT dataset.}\label{hyptab2}
\centering
\begin{tabular}{>{\centering\arraybackslash}m{2.8cm} >{\centering\arraybackslash}m{1.4cm} >{\centering\arraybackslash}m{1.4cm} >{\centering\arraybackslash}m{1.4cm} >{\centering\arraybackslash}m{1.4cm}}
\toprule
\multicolumn{1}{c} {Weight Scheme $^\star$} & MSE ($\downarrow$) & PSNR ($\uparrow$) & LPIPS ($\downarrow$) \\
\midrule
\multirow{1}{2.8cm}{[1, 1, 1, $10^{-2}$]} &0.00098&30.261&\underline{0.051}
\\
\multirow{1}{2.8cm}{[1, 1, 0, $10^{-2}$]}&\underline{0.00072}&\underline{34.167}&0.103
\\
\multirow{1}{2.8cm}{[1, 1, $10^{-1}$, $10^{-2}$]} &0.00090&30.302&0.054
\\
\multirow{1}{2.8cm}{[1, 1, $10^{-2}$, $10^{-2}$]} &0.00078&33.276&0.059
\\
\multirow{1}{2.8cm}{\textbf{[1, 1, \boldmath{$10^{-3}$}, \boldmath{$10^{-2}$}]}} & \textbf{0.00071} & \textbf{34.312} & \textbf{0.049}\\
\multirow{1}{2.8cm}{[1, 1, $10^{-3}$, $10^{-1}$]} &0.00073&33.060&0.113
\\
\multirow{1}{2.8cm}{[1, 1, $10^{-3}$, 0]} &0.00073&33.714&0.119
\\
\multirow{1}{2.8cm}{[1, 1, $10^{-3}$, 1]}&0.00101&31.427&0.137
\\
\multirow{1}{2.8cm}{[1, 0, $10^{-3}$, $10^{-2}$]} &0.00237&27.636&0.172
\\
\multirow{1}{2.8cm}{[0, 1, $10^{-3}$, $10^{-2}$]} &0.00151&28.392&0.059
\\
\bottomrule
\multicolumn{4}{l}{\small $\star$ $\lambda_{L1}$, $\lambda_{Qua}$, $\lambda_{Per}$ and $\lambda_{Adv}$, from left to right.}\\
\end{tabular}
\label{hyp2}
\vspace{-0.5cm}

\end{table}

\begin{table*}[!thbp]
\caption{Image quality assessment of BS-LDM scored by three radiologists with various levels of experience.}

\centering
\begin{tabular}{>{\raggedright\arraybackslash}p{3.4cm} >{\raggedright\arraybackslash}m{4.5cm} >{\centering\arraybackslash}m{2.4cm} >{\centering\arraybackslash}m{3.0cm} >{\centering\arraybackslash}m{2.4cm}}
\toprule
\multicolumn{2}{c} {Clinical Evaluation Criteria} & Junior Radiologist (6 years) & Intermediate Radiologist (11 years) & Senior Radiologist (21 years) \\
\midrule
\multirow{3}{3cm}{Lung vessel visibility} & Clearly displayed (3) & \multirow{3}{*}{} & \multirow{3}{*}{} & \multirow{3}{*}{} \\
 & Displayed (2) & 2.431 & 2.858 & 2.984 \\
 & Not displayed (1) &  &  &  \\
\midrule
\multirow{3}{3cm}{Airway visibility} & Lobar and intermediate bronchi (3) & \multirow{3}{*}{} & \multirow{3}{*}{} & \multirow{3}{*}{} \\
 & Main bronchus and rump (2) & 2.561 & 2.643 & 2.937 \\
 & Trachea (1) &  &  &  \\
\midrule
\multirow{3}{3cm}{Degree of bone suppression} & Nearly perfect suppression (3) & \multirow{3}{*}{} & \multirow{3}{*}{} & \multirow{3}{*}{} \\
 & Unsuppressed bones less than 5 (2) & 2.781 & 2.793 & 2.722 \\
 & 5 or more bones unsuppressed (1) &  &  &  \\
\bottomrule
\end{tabular}
\label{ima}
\vspace{-0.3cm}

\end{table*}

\subsection{Ablation Study}
To evaluate the importance of offset noise and the temporal adaptive thresholding strategy within the BS-LDM framework, we conducted experiments by training the model with and without these components on SZCH-X-Rays and JSRT datasets. When offset noise was absent, we applied Gaussian noise for evaluation. When temporal adaptive thresholding was absent, we applied static thresholding, dynamic thresholding, and no thresholding to evaluate their effects. Our results indicate that incorporating both offset noise and temporal adaptive thresholding is crucial for producing accurate soft tissue images, as it allows for better adjustment of low-frequency information. The quantitative data supporting these findings are presented in Table \ref{abl}, and the visual results can be seen in Fig. \ref{abl_img}.

\subsubsection{The Effect of Offset Noise}
In examining the effectiveness of offset noise, we found that its application resulted in generated soft tissue images with mean values more aligned with real images. This adjustment led to a reduction in MSE by at least 92.5\%, an improvement in PSNR by at least 0.897$dB$, and a decrease in LPIPS by at least 3.8\% compared to images generated without offset noise, across both datasets. Fig. \ref{abl_img} highlights that without offset noise, the brightness of the generated images was significantly misaligned.

\subsubsection{The Effect of Temporal Adaptive Thresholding}
Regarding temporal adaptive thresholding, expanding the thresholding range during sampling allowed for more precise adjustments of pixel intensities, as shown in Fig. \ref{abl_img}. Numerically, this method improved image contrast consistency with real images, achieving a reduction in MSE by at least 92.5\%, a PSNR increase of at least 0.897 $dB$, and a decrease in LPIPS by at least 3.8\% compared to other thresholding strategies or the absence of thresholding. In contrast, Fig. \ref{abl_img} illustrates that static and dynamic thresholding severely limited image variance, while the lack of thresholding resulted in excessive contrast due to pixel saturation issues.

\begin{figure}[!tbp]
\centering
\includegraphics[width=0.9\columnwidth]{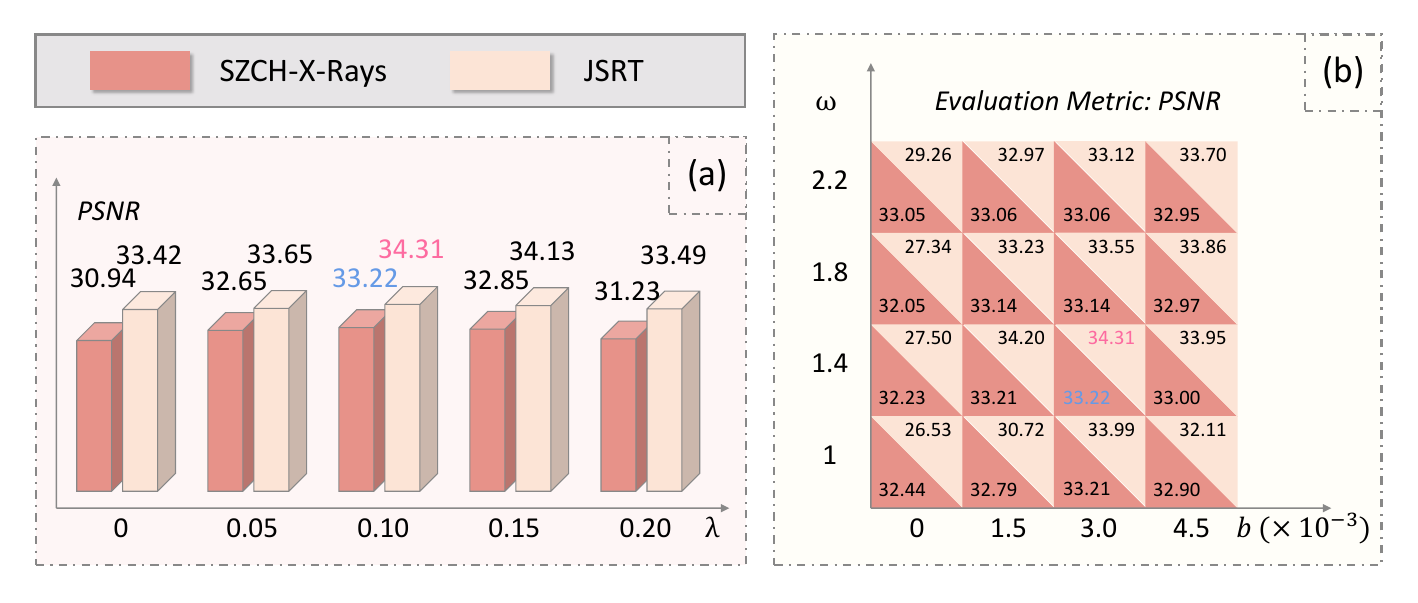}
\caption{Visualization of hyperparameter analysis of offset noise in (a) and the temporal adaptive thresholding strategy in (b) on SZCH-X-Rays and JSRT datasets.} \label{hyp}
\end{figure}

\begin{table}[!tb]
\caption{Diagnostic utility assessment of BS-LDM calculated from the diagnostic results of two radiologists with 6 and 11 years of experience, respectively.}\label{dia}
\centering

\begin{tabular}{p{2.4cm}<{\centering}p{1.5cm}<{\centering}p{1.8cm}<{\centering}p{1.5cm}<{\centering}}
\toprule
Junior Radiologist &  Precision ($\uparrow$) & Sensitivity ($\uparrow$) & F1 Score ($\uparrow$)\\
\midrule
CXR &  0.70 & 0.40 & 0.51\\
\textbf{Tissue} &  \textbf{0.73} & \textbf{0.56} & \textbf{0.63}\\
\midrule
Senior Radiologist &  Precision ($\uparrow$) & Sensitivity ($\uparrow$) & F1 Score ($\uparrow$)\\
\midrule
CXR &  0.74 & 0.51 & 0.60\\
\textbf{Tissue} &  \textbf{0.75} & \textbf{0.75} & \textbf{0.75}\\
\bottomrule
\end{tabular}
\vspace{-0.5cm}

\end{table}

\begin{table*}[!thbp]
\caption{Automated downstream evaluation of BS-LDM on Shenzhen chest X-ray using three classification models.}\label{aut}
\centering
\begin{tabular}{p{3cm}<{\centering}p{2cm}<{\centering}p{2cm}<{\centering}p{2cm}<{\centering}p{2cm}<{\centering}p{2cm}<{\centering}p{2cm}<{\centering}}
\toprule
Modal & Method &  Precision ($\uparrow$) & Sensitivity ($\uparrow$) & F1 Score ($\uparrow$) & Accuracy ($\uparrow$)& Specificity ($\uparrow$)\\
\midrule
\multirow{3}{*}{CXR}
&AlexNet \cite{krizhevsky2012imagenet}  &  0.816 $\pm$ 0.004 & 0.838 $\pm$ 0.068 & 0.826 $\pm$ 0.028& 0.821 $\pm$ 0.026& 0.803 $\pm$ 0.017\\
&DenseNet \cite{huang2017densely}  &  0.824 $\pm$ 0.014 & 0.926 $\pm$ 0.051 & 0.872 $\pm$ 0.032& 0.862 $\pm$ 0.026& 0.795 $\pm$ 0.035\\
&ResNet \cite{he2016deep} &  0.816 $\pm$ 0.077 &  0.912$\pm$ 0.051 & 0.861 $\pm$ 0.065& 0.851 $\pm$ 0.069& 0.788 $\pm$ 0.087\\
\midrule
\multirow{3}{*}{Soft Tissue}
&AlexNet \cite{krizhevsky2012imagenet} &  \textbf{0.855 $\pm$ 0.015} & \textbf{0.951 $\pm$ 0.017} & \textbf{0.898 $\pm$ 0.016}& \textbf{0.891 $\pm$ 0.017}& \textbf{0.828 $\pm$ 0.017}\\
&DenseNet \cite{huang2017densely} &  \textbf{0.833 $\pm$ 0.020} & \textbf{0.956 $\pm$ 0.051} & \textbf{0.890 $\pm$ 0.033} & \textbf{0.881 $\pm$ 0.034}& \textbf{0.803 $\pm$ 0.017}\\
&ResNet \cite{he2016deep} &  \textbf{0.841 $\pm$ 0.035} & \textbf{0.941 $\pm$ 0.068} & \textbf{0.888 $\pm$ 0.050} & \textbf{0.881 $\pm$ 0.052} & \textbf{0.818 $\pm$ 0.035}\\
\bottomrule
\end{tabular}
\vspace{-0.5cm}

\end{table*}

\subsection{Hyperparameter Analysis}
\subsubsection{Hyperparameter Analysis on the ML-VQGAN Loss}
We conducted a detailed analysis for each dataset to further examine the effects of the multi-level hybrid loss function of ML-VQGAN under varying hyperparameter settings. The results, as depicted in Table \ref{hyp1} and Table \ref{hyp2}, reveal that the BS-LDM framework achieves optimal performance when the weight parameters $\lambda_{L1}$, $\lambda_{Qua}$, $\lambda_{Per}$, and $\lambda_{Adv}$ are set to 1, 1, 0.001, and 0.01, respectively. The removal of the \(\mathcal{L}_{Qua}\) results in a noticeable decline across all evaluation metrics. Even if the primary reconstruction loss is replaced with \(\mathcal{L}_{Per}\), the absence of the \(\mathcal{L}_{L1}\) leads to a significant reduction in pixel-level performance, though the impact on LPIPS is relatively minor. Moreover, when both the \(\mathcal{L}_{Per}\) and \(\mathcal{L}_{Adv}\) are omitted, the generated images exhibit a lack of texture detail. This deficiency is reflected in a consistent drop across all metrics, underscoring the importance of these components for achieving high-quality results.
\subsubsection{Importance of Parameters on Offset Noise and Temporal Adaptive Thresholding}
To further investigate the impact of offset noise and the temporal adaptive thresholding strategy, we conducted a hyperparameter analysis on each dataset. The results, shown in Fig. \ref{hyp}, suggest that optimal model performance is achieved when the bias noise weight \(\lambda\) in offset noise is set to 0.1, and the parameters \(\omega\) and \(b\) for temporal adaptive thresholding are set to 0.003 and 1.4, respectively. 
In offset noise, a large \(\lambda\) can disproportionately alter the model's behavior, whereas a small \(\lambda\) may render the effect negligible. Similarly, in the temporal adaptive thresholding strategy, the values of \(\omega\) and \(b\) are influenced by the data characteristics, and extreme values can lead to a decline in performance.

\subsection{Clinical Evaluation}
\subsubsection{Image Quality Assessment}
The soft tissue images generated by the BS-LDM on the SZCH-X-Rays dataset were independently evaluated for image quality using established clinical criteria \cite{bae2022bone, hong2021value} that are commonly applied to assess bone suppression efficacy. Three radiologists, with 6, 11, and 21 years of experience respectively, conducted these evaluations at our partner hospital. The detailed results are presented in Table \ref{ima}. The average scores for lung vessel visibility, airway visibility, and the degree of bone suppression were 2.758, 2.714, and 2.765, respectively, out of a maximum score of 3. These findings indicate that BS-LDM effectively suppresses bone while preserving fine details and lung pathology.

\subsubsection{Diagnostic Utility Assessment}
The diagnostic value of soft tissue imaging was independently evaluated by two radiologists with 6 and 11 years of experience, following the X-ray diagnosis standard \cite{bansal2019interpreting}. This analysis employed the SZCH-X-Rays dataset for bone suppression, using computed tomography to confirm lesions, which included common lung diseases such as inflammation, tuberculosis, and masses or nodules. Out of 818 data pairs assessed, 79 pairs contained one or more of these lesions. The radiologists independently evaluated both conventional CXR and the soft tissue images generated by our model. Table~\ref{dia} summarizes the evaluation results, focusing on key diagnostic metrics such as Precision, Sensitivity, and F1 Score. The findings reveal that the soft tissue images generated by BS-LDM significantly enhance diagnostic performance for both junior and senior radiologists compared to conventional CXR images. Notably, the diagnostic accuracy achieved by junior radiologists using the generated soft tissue images closely matches the performance of senior radiologists working with CXR images. These results underscore the potential of BS-LDM to improve lesion detection and diagnosis, offering a more comprehensive and accurate assessment. This highlights its substantial value in clinical applications, where it could serve as an effective tool to augment diagnostic capabilities across varying levels of expertise.

\subsection{Automated Downstream Evaluation}
To objectively evaluate the effectiveness of our method for lung disease diagnosis, we performed an automated downstream analysis on the widely used public Shenzhen chest X-ray dataset \cite{jaeger2014two}. This dataset includes 336 radiographs of tuberculosis cases and 326 normal cases. Using the BS-LDM model trained on SZCH-X-Rays, we generated corresponding soft tissue images for the Shenzhen chest X-ray dataset. Both the original CXR images and the generated soft tissue images were input into lung disease prediction systems based on AlexNet \cite{krizhevsky2012imagenet}, DenseNet \cite{huang2017densely}, and ResNet \cite{he2016deep} for evaluation. The quantitative results, presented in Table \ref{aut}, demonstrate consistent performance improvements across key metrics, including Precision, Sensitivity, F1 Score, Accuracy, and Specificity. Notably, when using soft tissue images generated by BS-LDM, Sensitivity improved by 13.5\%, 3.24\%, and 3.18\% in AlexNet, DenseNet, and ResNet models, respectively. These enhancements highlight the capability of our method to augment downstream diagnostic performance in CXR imaging, further supporting its clinical utility in improving lung disease detection.

\subsection{Limitations}
While BS-LDM has shown remarkable capability in producing high-resolution soft tissue images from CXR data, its application in clinical practice remains limited. To enable successful integration into existing Picture Archiving and Communication Systems (PACS), the following key challenges have to be addressed:
\begin{itemize}
\item Although BS-LDM has achieved notable improvements in inference efficiency, surpassing other diffusion-based methods by 28.6\%, as shown in Fig. \ref{bar}, there remains room for further optimization. To enhance its performance, we intend to investigate the incorporation of advanced denoising samplers, such as DPM \cite{lu2022dpm} and LCM \cite{luo2023latent}. These approaches have the potential to reduce the number of sampling steps, thereby increasing the model's overall speed and efficiency.
\item Processing high-resolution images often demands substantial memory, which may restrict the practical use of BS-LDM in clinical settings with limited computational resources. To overcome this challenge, we are investigating architectural advancements that aim to reduce memory consumption while preserving performance. Promising approaches involve incorporating Transformer models \cite{peebles2023scalable} or Structured Space Models \cite{fei2024scalable} as noise estimator networks. These innovations could enable a more memory-efficient architecture that retains the precise detail required for medical imaging, thereby enhancing the model's suitability for deployment across clinical environments with diverse resource constraints.
\item Ensuring compatibility with standard medical image formats, such as DICOM, is essential for the seamless integration of BS-LDM into PACS. To address this, we are evaluating BS-LDM's ability to handle DICOM-formatted data during both pre-processing and post-processing stages, thereby streamlining its incorporation into routine clinical workflows.
\item Developing a user-friendly interface is crucial for integrating BS-LDM seamlessly into existing clinical workflows. To ensure practicality and ease of use, we are collaborating with radiologists to design an intuitive interface that streamlines interaction with the model, promoting greater user adoption and enhancing its clinical utility.
\end{itemize}
In future research, we aim to address these limitations and advance the clinical application of our method.

\section{Conclusion}

To enhance radiologists' ability to detect lung lesions with greater precision and thoroughness, we introduce BS-LDM, an end-to-end framework for bone suppression in high-resolution CXR images, utilizing conditional LDMs. We utilize a multi-level hybrid loss-constrained VQGAN for perceptual compression to ensure detail preservation. To further refine the framework's performance, we incorporate offset noise and a temporal adaptive thresholding strategy. These enhancements help reduce discrepancies in generating low-frequency information, thereby enhancing the quality and interpretability of the soft tissue images produced. Our framework effectively generates soft tissue images that achieve high levels of bone suppression while retaining essential details and preserving the integrity of lung lesions. Extensive experimental and downstream evaluations validate the superior bone suppression performance of BS-LDM and underscore its substantial clinical value. Although BS-LDM demonstrates potential, its clinical application is hindered by several obstacles. Key challenges include improving inference efficiency, optimizing memory usage, ensuring compatibility with the DICOM format, and refining interface design. Addressing these issues will be the focus of future efforts to enhance its viability for use in healthcare environments.

\section{Compliance With Ethical Standards}

This research was conducted retrospectively and following the principles of the Declaration of Helsinki, utilizing previously obtained chest X-ray images that had been anonymized before conducting analysis, thus posing no harm to patients. We applied for an informed consent waiver. Approval was granted by the Ethics Committee of Cancer Hospital \& Shenzhen Hospital, Chinese Academy of Medical Sciences and Peking Union Medical College (No. JS2023-19-2).
\section*{References}

\bibliographystyle{IEEEtran}
\bibliography{ieee_bib}
\end{document}